\documentclass[fleqn,usenatbib]{mnras}
\usepackage{newtxtext,newtxmath}
\usepackage[T1]{fontenc}
\DeclareRobustCommand{\VAN}[3]{#2}
\let\VANthebibliography\thebibliography
\def\thebibliography{\DeclareRobustCommand{\VAN}[3]{##3}\VANthebibliography}
\usepackage{graphicx}	
\usepackage{amsmath}
\usepackage{lineno}
\usepackage{color}
\usepackage{graphicx}
\usepackage{longtable}
\usepackage{multirow}
\usepackage{url}
\usepackage{subfigure}
\usepackage{lipsum}
\usepackage{ulem}
\usepackage{bm}
\usepackage{threeparttable}
\usepackage{orcidlink}

\title[FRB-like bursts, glitch and anti-glitch]{Tidal capture of an asteroid by a magnetar: FRB-like bursts, glitch and anti-glitch}

\author[Wu, Zhao \& Wang]{
Qin Wu\orcidlink{0000-0001-6021-5933},$^{1}$
Zhen-Yin Zhao\orcidlink{0000-0002-2171-9861},$^{1}$
Fa-Yin Wang\orcidlink{0000-0003-4157-7714}$^{1,2,3}$ 
\thanks{E-mail: fayinwang@nju.edu.cn}
\\
$^{1}$School of Astronomy and Space Science, Nanjing University, Nanjing 210093, China\\
$^{2}$Key Laboratory of Modern Astronomy and Astrophysics (Nanjing University), Ministry of Education, Nanjing 210093, China\\
$^3$ Purple Mountain Observatory, Chinese Academy of Sciences, Nanjing, 210023, China}

\date{Accepted XXX. Received YYY; in original form ZZZ}
\pubyear{2023}

\begin{document}
\label{firstpage}
\pagerange{\pageref{firstpage}--\pageref{lastpage}}
\maketitle

\begin{abstract}
Recently, remarkable anti-glitch and glitch accompanied by bright radio bursts of the Galactic magnetar SGR J1935+2154 were discovered. 
These two infrequent temporal coincidences between the glitch/anti-glitch and the fast radio burst (FRB)-like bursts reveal their physical connection of them. 
Here we propose that the anti-glitch/glitch and FRB-like bursts can be well understood by an asteroid tidally captured by a magnetar. 
In this model, an asteroid is tidally captured and disrupted by a magnetar.
Then, the disrupted asteroid will transfer the angular momentum to the magnetar producing a sudden change in the magnetar rotational frequency at the magnetosphere radius. 
If the orbital angular momentum of the asteroid is parallel (or anti-parallel) to that of the spinning magnetar, a glitch (or anti-glitch) will occur. 
Subsequently, the bound asteroid materials fall back to the pericenter and eventually are accreted to the surface of the magnetar. 
Massive fragments of the asteroid cross magnetic field lines and produce bright radio bursts through coherent curvature radiation. 
Our model can explain the sudden magnetar spin changes and FRB-like bursts in a unified way.
\end{abstract}


\section{Introduction}
SGR J1935+2154 is a Galactic magnetar with a spin period of $P\sim 3.24 \ \rm{s}$ and a magnetic field strength of $B\sim 2.2\times 10^{14}\ \rm{G}$ \citep{Israel2016}.
On 28 April 2020, a bright millisecond radio burst FRB 20200428 from SGR J1935+2154 was detected by CHIME and STARE2 \citep{CHIME2020, Bochenek2020}.  
Meanwhile, a bright X-ray burst was temporally coincident with FRB 20200428 \citep{Mereghetti2020, Li2021, Ridnaia2021, Tavani2021}.
The detection of FRB 20200428 from a Galactic magnetar indicates that at least part of FRBs originates from magnetars \citep{WangF2022}.
Some models have been proposed to explicate the simultaneous production of FRB 20200428 and X-ray burst from SGR J1935+2154 \citep{Dai2020, Geng2020, Wu2020, Margalit2020, Lu2020, Shen2023}.

From 1 October to 27 November 2020, the NICER and XMM-Newton telescopes monitored SGR J1935+2154 regularly in the 1-3 keV band.
A timing analysis of the X-ray pulses was employed and an anti-glitch with $\Delta\nu = 1.8^{+0.7}_{-0.5}\times 10^{-6}\ \rm{Hz}$ was found to occur on 5 October 2020 (+/-1 day) \citep{Younes2022, Zhu2020}. 
Subsequently, three bright radio bursts were detected by CHIME on 8 October 2020 \citep{Good2020}.
The fluence of three FRB-like bursts is 900$\pm$160 Jy ms, 9.2$\pm$1.6 Jy ms, and 6.4$\pm$1.1 Jy ms, respectively. 
Considering the distance of SGR J1935+2154 is about $D = 9\ \rm{kpc}$ \citep{Zhong2020}, the energies of radio bursts are $3.5\times 10^{31}$ erg, $3.6\times 10^{29}$ erg and $2.5 \times 10^{29}$ erg.
More recently, \cite{Ge2022} studied the timing properties of SGR J1935+2154 using the archived NICER, NuSTAR, Chandra, and XMM-Newton observations, and reported that a giant spin-up glitch with $\Delta\nu = 1.98\pm 0.14 \times 10^{-6}\ \rm{Hz}$ happened $3.1\pm 2.5$ days before FRB 20200428. 

Glitch is a frequent phenomenon observed in many pulsars, which is defined as a sudden increase in rotational frequency. The origin of glitches is still unknown. They can be caused by internal processes within the pulsar or external processes outside the pulsar \citep{Haskell2015}. 
Anti-glitch is characterized by a sudden decrease in rotational frequency, which is not common as a glitch. 
Some models have been proposed to explain the origin of anti-glitches \citep{Thompson2000, Lyutikov2013, Tong2014, Katz2014, Huang2014, Kantor2014}. 
The accetion of planetesimals by a magnetar and the collision of a solid body with a magnetar was used to explain the anti-glitch from magnetar 1E 2259+586 \citep{Katz2014, Huang2014}. 

Different from normal glitches and anti-glitches of pulsars, the glitch, and anti-glitch of SGR J1935+2154 are accompanied by FRB-like bursts.
The temporal coincidence between the glitch/anti-glitch and the FRB-like bursts of SGR J1935+2154 indicate a physical connection between them \citep{Younes2022, Ge2022}. 
\cite{Younes2022} proposed that the rapid reduction of the angular momentum of magnetar comes from the escape of the particle wind along the open magnetic field lines. While the subsequent radio bursts arise from the alteration of the magnetospheric field geometry.
\cite{Ge2022} supplied an interpretation that the spin-up glitch before FRB 20200428 is produced by the violent movement of the magnetosphere. While the crustal cracking and the propagation of Alfv\'{e}n wave produced the following FRB 20200428 and the associated X-ray bursts. 
There are also other discussions on the glitch \citep{WangWH2022}, the anti-glitch \citep{WangWH2021, Tong2022}, and the glitch-associated activities \citep{Du2023} of SGR J1935+2154. 
However, neither of them provides a self-consistent model that can explain the anti-glitch/glitch and FRB-like bursts simultaneously. 

In this paper, we propose that an asteroid captured by a magnetar (hereafter the capture model) can explain the glitch and anti-glitch of SGR J1935+2154 and the subsequent FRB-like bursts simultaneously.
The capture model and the parameter constraints are given in Section \ref{sec2}.
In the last, the discussion and conclusion are shown in Section \ref{sec3}.

\begin{figure*}
	\centering
	\includegraphics[width=\linewidth]{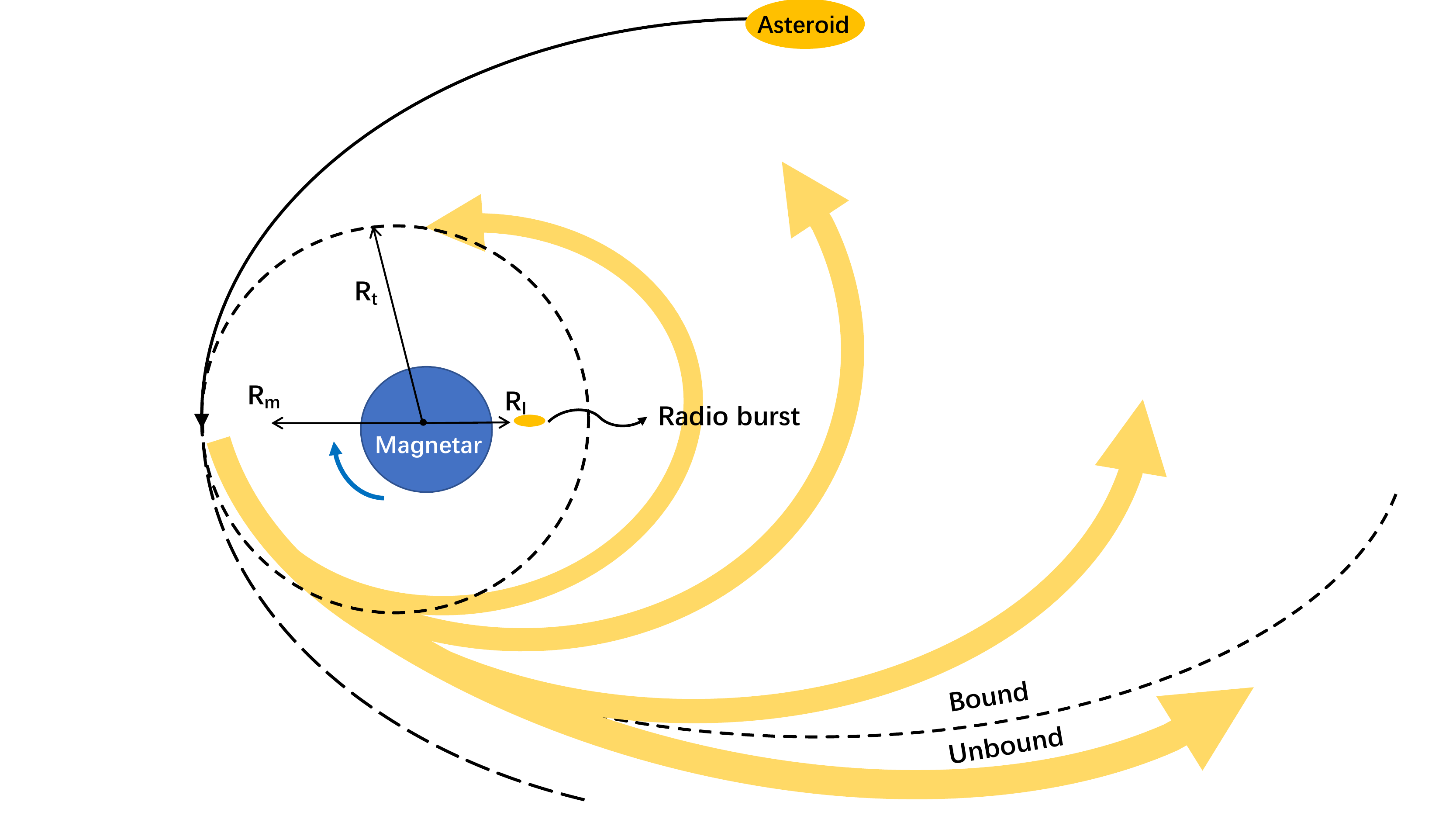}
	\caption{\label{fig} Schematic diagram of the capture model.
		(a) An iron-nickel asteroid with mass $m$ is tidally captured and disrupted by a magnetar at the tidal disruption radius $R_{\rm t}$. 
		(b) The orbital angular momentum of the asteroid is transferred to the magnetar at the magnetosphere radius $R_{\rm m}$, which will cause the observed anti-glitch or glitch. The blue arrow represents the direction of the spin of the magnetar. 
		A fraction of the disrupted material will escape the magnetar at high speed, while the rest will be bound to the magnetar and experience a process of orbital circularization. The bound material will fall back to the pericenter and eventually be accreted to the magnetar. 
		(c) The broken fragment with the mass of $m_0$ crosses the magnetic field lines and induces the FRB-like burst through coherent curvature radiation at the interaction radius $R_{\rm I}$. This capture model can explain the glitch/anti-glitch and the associated FRB-like bursts in a unified way.}
\end{figure*}

\section{The Capture Model}\label{sec2}
We consider that an asteroid is tidally captured and disrupted by the strong gravitational force of a magnetar when it approaches the tidal disruption radius (see Figure \ref{fig}). 
A part of the disrupted material will be accreted and experience orbital circularization, while the rest material will be thrown out at a high speed. 
The broken and accreted asteroid transfers the angular momentum to the magnetar at the magnetosphere radius, which causes a sudden change in the rotational frequency of the magnetar \citep{Ghosh1979a}. 
When the orbital angular momentum of the asteroid is parallel to the spin of the magnetar, an increase in the rotational frequency naturally happens. 
On the contrary, the anti-parallel orientation will cause a decrease in the rotational frequency of the magnetar. 
Then the bound asteroid will fall back to the magnetar due to dissipative processes such as collision and viscous dissipation and eventually be accreted onto the surface of the magnetar \citep{Rees1988, Perets2016}. 
The FRB-like bursts are naturally produced when the broken fragments are across the magnetic field lines through coherent curvature radiation processes \citep{Dai2016, Dai2020}.

\subsection{Glitch and anti-glitch}

Here we suppose an asteroid is disrupted tidally by the magnetar with typical mass $M_{\rm NS} = {\rm 1.4\ M_{\odot}}$ and radius $R_{\rm NS} = 10$ km at the tidal disruption radius \citep{Hills1975}
\begin{equation}\label{disrupt radius}
    R_{\mathrm{t}} \approx(6 M_{\rm NS} / \pi \rho)^{1 / 3} 
    \sim 8.7 \times 10^{10}(M_{\rm{NS}}/1.4M_{\odot})^{1/3}\rho_{0.9}^{-1/3}\ \rm{cm},
\end{equation}
where the expression $Q_{x}$ represents $Q/10^{x}$ in cgs unit system and $\rho =  {\rm 8\ g\ cm^{-3}} $ is the density of a homogeneous iron-nickel asteroid \citep{Colgate1981}. 
Then a fraction of the disrupted material is tidally captured by the magnetar while the rest will be thrown out. 
The bound material will enter an elliptical orbit, which length of the semi-major axis can be estimated as ${R_{\rm orb}\sim4.4\times10^{14}}$ cm \citep{Zubovas2012}. The orbital period is calculated as 135 years using Kepler's third law. After a few orbits, the circularization can occur \citep{Cannizzo1990}. The material will fall back to the magnetar vicinity within this timescale \citep{Rees1988}.
The ram pressure is generated during the falling of the bound material.  
Considering the balance of the ram pressure and the magnetic pressure, we can calculate the magnetosphere radius $R_{\rm m}$ of the disrupted material \citep{Ghosh1979a}
\begin{equation}\label{Magnetosphere radius}
    R_{\mathrm{m}}=\left(\frac{\mu^4}{2 G M_{\rm NS} \dot{m}^2}\right)^{1 / 7},
\end{equation}
where $\mu = B R_{\rm NS}^{3}$ is the magnetic dipole moment and the surface magnetic filed is $B = 3.52\times 10^{14} \ G$ for SGR J1935+2154.

From the magnetosphere radius $R_{\rm m}$, the orbital angular momentum of the asteroid will transfer to the magnetar, which will cause the observed glitch or anti-glitch. 
The velocity of the asteroid with mass $m$ is $V_m = (2GM_{\rm NS}/R_{\rm m})^{1/2}$ and the orbital angular momentum of the asteroid can be calculated as $-m V_{\rm m} R_{\rm m}$.  
Conservation of angular momentum for the system composed of a magnetar and an asteroid can be written as 
\begin{equation}\label{moment}
    2 \pi \nu I - m V_{\rm m} R_{\rm m} = 2 \pi (\nu - \Delta \nu) I,
\end{equation}
where $I$ is the moment of inertia of the magnetar and $\Delta \nu$ is the change of frequency. 
Considering a magnetar with mass $M_{\rm NS} = 1.4M_{\odot}$ and radius $R_{\rm NS} = 10$ km, the moment of inertia of the magnetar $I = \frac{2}{5}M_{\rm NS}R_{\rm NS}^2= 1.12\times 10^{45} \ {\rm g}\ {\rm cm}^2$ \citep{Pizzochero2011, Hooker2015}.

In our model, the disrupted material is assumed to be accreted and adheres to the central magnetar from the magnetosphere radius \citep{Ghosh1979a}. 
The average mass accretion rate is 
\begin{equation}\label{accretion}
    \dot{m} = \frac{m_{\rm acc}}{\Delta t},
\end{equation}
where $m_{\rm acc}$ is the mass of accreted material and $\Delta t$ is the accretion timescale, which is the time from the initial accretion of the asteroid to the magnetosphere radius. 
The value of ${\Delta t}$ is hard to determine from observations. Here we suppose that the time of accretion is about ${\Delta t\sim 1}$ hour, which is shorter than the time lag between the glitch/anti-glitch and the radio burst \citep{Younes2022, Ge2022}.

Utilizing equations (\ref{Magnetosphere radius}), (\ref{moment}) and (\ref{accretion}), we can derive the mass accretion rate $\dot{m}$.
For the anti-glitch with $\Delta\nu_1 = 1.8^{+0.7}_{-0.5}\times 10^{-6}$ Hz \citep{Younes2022}, the mass accretion rate $\dot{m_1}$ is given 
\begin{equation}
\begin{split}
    \dot{m_1} 
    &= \frac{(2\pi I \Delta\nu_1)^{7/6}}{\mu^{1/3}(2GM_{\rm NS})^{1/2} \Delta t^{7/6}}  \\
    &\sim {7.82\times10^{16}}I_{45}^{7/6}\Delta\nu_{1,-5.7}^{7/6}\mu_{32.5}^{-1/3}\Delta t_{4.9}^{-7/6} \ \rm{g \ s^{-1}}.
\end{split}
\end{equation}
Then the accretion mass of the anti-glitch $m_1$ is 
\begin{equation}
    m_1 = \dot{m_1} \Delta t  
    = {6.76\times10^{21}}I_{45}^{7/6}\Delta\nu_{1,-5.7}^{7/6}\mu_{32.5}^{-1/3}\Delta t_{4.9}^{-1/6}\ \rm{g},
\end{equation}
which is the lower limit of the mass of the asteroid.
The magnetosphere radius $R_{\rm m1}$ of the anti-glitch is estimated as 
\begin{equation}
    R_{\rm m1} = \left(\frac{\mu^2\Delta t}{2\pi I_c\Delta \nu_1}\right)^{1/3} =  {9.45\times10^{9}}\mu_{32.5}^{2/3}\Delta t_{4.9}^{1/3}I_{45}^{-1/3}\Delta\nu_{1,-5.7}^{-1/3} \ \rm{cm},
\end{equation}
which is smaller than the tidal disruption radius $R_{\rm t}$ (equation (\ref{disrupt radius})). 

If the orbital angular momentum of the asteroid is parallel to the spin of the magnetar, a spin-up glitch naturally occurs. 
For the spin-up glitch with ${\Delta\nu_2 = 1.98\times 10^{-5}}$ Hz \citep{Ge2022}, the mass accretion rate is ${\dot{m_2} \simeq 1.28\times 10^{18}}\ \rm{{g\ s^{-1}}}$, the accretion mass is ${m_2\simeq 1.11\times 10^{23}}$ g and the corresponding magnetosphere radius is ${R_{\rm m2}\simeq 4.25\times 10^{9}}$ cm.

The radius of the light-cylinder of SGR J1935+2154 is calculated as
\begin{equation}
    {R_{\rm L}= \frac{c}{2\pi\nu}=1.55\times10^{10}}\ \rm{ cm},
\end{equation}
where ${\nu=0.307896}$ Hz is the spin frequency of SGR J1935+2154. The magnetic field lines are connected to the magnetar inside the light cylinder. The magnetosphere radius ${R_{\rm{m}}}$ for the glitch/anti-glitch is well inside the light cylinder. So the angular momentum of the asteroid can be transferred to the magnetar.

After the tidal disruption, a fraction of the disrupted and bound debris may fall back and self-interact. 
Then the orbiting debris with high eccentricity will experience orbital circularization and a transient disk is formed around the magnetar. 
As the corotation radius ${R_{\rm co}\sim4.6\times10^8}$ cm is smaller than the magnetosphere radius ${R_{\rm{m}}}$, the bound material will accumulate at the magnetosphere radius and eventually fall back to the magnetar once the ram pressure exceeds the magnetic pressure. The glitch/anti-glitch happens due to the angular momentum transfer between the bound material and the magnetar at the magnetosphere radius ${R_{\rm{m}}}$. Then the bound material will fall back to the vicinity of the magnetar and produce bright radio bursts.  
The fallback timescale that the bound asteroid from the magnetosphere radius returns to the vicinity of the magnetar can be calculated as \citep{Perets2016}
\begin{equation}
    {t_{\mathrm{fb}}=\frac{2 \pi R_{\mathrm{m}}^3}{(G M_{\rm NS})^{1 / 2}(2 r)^{3 / 2}}\sim 3.6\ r_{6.8}^{-3/2}\mu_{32.5}^{2}\Delta t_{4.9}I_{45}^{-1}\Delta\nu_{1,-5.7}^{-1}}\ \rm{{days}},
\end{equation}
where ${r}$ is the radius of the asteroid. 
The fallback time is comparable to the observed time delay ($\sim$ 3 days) between the glitch/anti-glitch and the FRB-like bursts \citep{Younes2022, Ge2022}, which means our model is self-consistent. 
The bound disk formed by the fallback material will be accreted by the magnetar and produce bright radio bursts when the large fragments cross the magnetic lines. 
\cite{Dai2020} gave an appropriate explanation for the production of FRB 20200428 and the associated X-ray burst, which are temporal coincident with the spin-up glitch \citep{Ge2022}. 
In the following section, we discuss the generation of the FRB-like bursts that occurred on 8 October 2020 after the anti-glitch \citep{Good2020, Younes2022}.

\subsection{FRB-like bursts}

The asteroid is firstly tidally disrupted into a great number of fragments at the tidal disruption radius $R_{\rm t}$.  
The broken fragments will slow down in a strong magnetic field of the magnetar. 
Three massive fragments are distorted by the magnetar at the broken-up radius $R_{b} = (\rho r_0^2GM_{\rm NS}/s)^{1/3}$ with the tensile strength $s$ and the original radius $r_0$ of the fragment \citep{Colgate1981}. 
The broken fragment will be elongated longitudinally as a cylinder radius $r_0$ and length $l_0 = 2r_0$ and will be further disrupted into a train of small chunks. Bright FRB-like bursts are produced when these disrupted fragments cross the magnetic lines through curvature radiation. 
The time difference between the leading and lagging parts of a train of small fragments reaching the surface of the magnetar is  
\begin{equation}\label{time difference}
    \delta t \simeq \frac{12 r_{0}}{5}\left(\frac{2 G M_{\rm NS}}{R_{b}}\right)^{-1 / 2},
\end{equation}
where $s \sim 10^{10}\ \rm{dyn\ cm^{-2}}$ for the iron-nickel asteroid. 
This time difference is comparable to the duration of an FRB-like burst, which is independent of the position of the fragments \citep{Geng2015}.
The duration of the first and the brightest burst of three FRB-like bursts is $\delta t \sim 0.26\pm 0.01$ ms \citep{Good2020}, thus the collective mass of a train of fragments $m_0$ can be calculated as 
\begin{equation}\label{fragment mass}
\begin{split}
    m_0 =  2\pi\left(\frac{5\delta t}{12}\right)^3\left(\frac{2GM_{\rm NS}}{R_b}\right)^{3/2}  = 1.7\times 10^{16}\delta t_{-3.6}^{3} \rho_{0.9} R_{\rm b,9.8}^{-3/2} \  \rm{g}. 
\end{split}
\end{equation}
The break-up radius $R_b \sim 5.9\times 10^9\ \rm{cm}$ is smaller than the tidal disruption radius $R_{\rm t}\sim 8.7\times 10^{10}\ \rm{cm}$ and the magnetosphere radius of the anti-glitch ${R_{\rm m1}\sim9.45\times10^{9}}\ \rm{cm}$, implying that our calculations are self-consistent. 

The disrupted fragments are apparently influenced by the strong magnetic field of the magnetar at the interaction radius ($R_{\rm I}$) where the kinetic energy density of the fragment $\rho v_{\rm ff}^2/2$ is equal to the magnetic energy density $\mu^2/ 8\pi R_{\rm I}^6$. 
Here we give the interaction radius $R_{\rm I}$ of the broken-up fragments
\begin{equation}
    R_{\rm I} = 3.05\times 10^7\rho_{0.9}^{-1/5}B_{14.5}^{2/5}R_{\rm NS, 6}^{6/5} \ \rm{cm},
\end{equation}
where $v_{\rm ff} = \sqrt{2GM_{\rm NS}/R_{\rm I}}$ and the magnetic dipole moment of the magnetar $\mu = BR_{\rm NS}^3$.
The disrupted bound material will induce a strong electric field (${E} = -v_{\rm ff}\times B / c$) when it penetrates through the stellar magnetosphere \citep{Lamb1973, Burnard1983, Hameury1986, Litwin2001}. 
The electrons are accelerated to ultra-relativistic energies and move along magnetic field lines. 
Movement of electrons along magnetic field line with curvature radius $\rho_c$ generates coherent curvature radiation and leads to FRB-like bursts at the radius $R_{\rm I}$ \citep{Dai2016, Dai2020}. 
The typical Lorentz factor $\gamma$ of an
accelerated electron is given by
\begin{equation}
    \gamma \equiv \chi\gamma_{\rm max} \simeq \chi\left(\frac{6\pi e E}{\sigma_{\rm T}B^2}\right)^{1/2} = 216.07 \chi R_{\rm I, 7.5}^{5/4}\mu_{32.5}^{-1/2} , 
\end{equation}
where $\sigma_{\rm T}$ is the Thomson scattering cross section and $\chi$ is defined as the ratio of $\gamma$ and the maximum Lorentz factor $\gamma_{\rm max}$ \citep{Dai2016}. 
Condition $\chi\lesssim 1$ needs to be satisfied and we generally assume $\chi \simeq 0.5$, then $\gamma \sim 108$. 

The characteristic frequency of the curvature radiation is 
\begin{equation}
    \nu_{\rm c} = \frac{3c\Gamma\gamma^{3}}{4\pi \rho_c} = 465.49 \delta R_{\rm I, 7.5}^{11/4}\mu_{32.5}^{-3/2}    \ \rm{MHz},
\end{equation}
where curvature radius $\rho_c\sim 0.635 R_{\rm I}$ is assumed \citep{Yang2018} and ${\Gamma}$ is the factor related to the Doppler effect \citep{Dai2020}. 
This characteristic frequency is consistent with the FRB-like bursts detected by CHIME in the range 400-600 MHz \citep{Good2020}. 

The disrupted and bound fragments are radially elongated and transversely compressed to a cylindrical shape from the break-up radius ${R_{\rm b}}$. Some small fragments will undergo further disruption into smaller chunks and move with the closed magnetic field lines \citep{Metzger2018}. A train of chunks will cross the magnetic field lines and produce the bright radio bursts through the coherent curvature radiation at the interaction radius ${R_{\rm I}}$. And an openmouthed-clam-shaped radiating region is naturally formed with an inclination angle of ${\theta_{\rm i}\simeq r(R_{\rm I})/R_{\rm I}}$ \citep{Dai2020}. 
The total luminosity of an FRB-like burst is derived by considering an openmouthed-clam-shaped radiating region \citep{Dai2020}, 
\begin{equation}
    L_{\rm tot} = 3.32\times 10^{32} R_{\rm I, 7.5}^{-27/4}\mu_{32.5}^{3/2} m_{0, 16.2}^{8/9}\rho_{0.9}^{-14/9}s_{10}^{2/3}       \ \rm{erg\ s^{-1}}. 
\end{equation}
In the observer's rest frame, the isotropic-equivalent emission energy of the FRB-like burst is given by 
\begin{equation}
    E_{\rm radio} \simeq \frac{L_{\rm tot}\delta t\Gamma^3}{f}\sim 1.86\times 10^{31}\Gamma^3  \ \rm{erg}, 
\end{equation}
where $f\simeq 1/2\gamma$ is the beaming factor and ${\delta t\sim 0.26\pm0.01}$ ms is the duration of the radio burst. 
This is comparable to the energy of three FRB-like bursts observed by CHIME on 8 October \citep{Good2020}, which indicates the FRB-like burst can be naturally produced in our model. 

When the asteroid gets closer to the magnetar, evaporation caused by the heating of the asteroid should be considered. Considering the thermal equilibrium with the magnetar surface radiation, the temperature of the asteroid is ${T_{\rm e}=T_{\rm sur}\sqrt{R_{\rm NS}/2r}}$ \citep{Cordes2008}. Here we assume that the surface temperature of the asteroid is ${T_{\rm sur}\sim10^5}$ K regarding the age and the mass of this relatively old magnetar. At the broken-up radius ${R_{\rm b}\sim5.9\times10^9}$ cm, the temperature of the asteroid is ${T_{\rm e}\sim920}$ K, which is smaller than the evaporation temperature of the iron ($\sim$ 2000 K). Thus, the curvature radiation will not be significantly affected by evaporation before the tidal break-up. 
Heating by high-energy particles by the non-thermal radiation from the magnetar will not significantly evaporate the asteroid, because of the decreasing kinetic energy of the high-energy particles when the bound material approaches the magnetar \citep{Geng2020}. However, the evaporation can be significant for the small fragment because the evaporation happens faster for the smaller asteroid \citep{Zubovas2012}. The study on the evaporation near supermassive black holes had been performed by \cite{Zubovas2012}.

\section{Discussion and Conclusions}\label{sec3}
Observations show that X-ray bursts are temporal coincident with FRB 20200428 about 3 days after the glitch. \cite{Ge2022} proposed that FRB 20200428 and the associated X-ray bursts originate from the violent alteration of magnetosphere structure. 
However, due to the lack of model details, whether it can explain the observation of the anti-glitch is unclear. \cite{Younes2022} postulated that an ephemeral and strong wind that extracts angular momentum from the magnetar leads to an abrupt spin-down of SGR J1935+2154. 
Obviously, the outward wind is difficult to explain the sudden spin-up of SGR J1935+2154 \citep{Ge2022}.
The energy ratio between FRB 20200428 and the X-ray burst from SGR J1935+2154 is $\eta = 2.13\times 10^{-5}$ \citep{Wu2020}.  
If a similar ratio is assumed for the radio bursts observed in October 2020,   
the luminosity of the X-ray burst associated with the FRB-like burst can be estimated as 
\begin{equation}
    L_{\rm X} =  \frac{L_{\rm radio}}{\eta}\sim 8.74 \times 10^{35}
    \eta_{-4.7}^{-1} L_{\rm tot, 32.5}\ \rm{erg\ s^{-1}}.
\end{equation}
This is comparable to the 5$\sigma$ detection upper-limit of the X-ray luminosity, i.e., $\rm 10^{36}\ erg\ s^{-1}$ \citep{Younes2022}. 
Thus the absence of X-ray bursts after the anti-glitch can be naturally explained as being too faint to be observed.

The temporal coincidence between the infrequent glitch/anti-glitch and bright radio bursts implies some physical association between them, which
is also supported by statistical similarities.
The size distribution of glitches in individual pulsars is power-law form \citep{Morley1993, Melatos2008}. The waiting times of glitches also show power-law distribution \cite{Carlin2021}, which is consistent with a non-stationary Poisson process.
Power-law distributions of energy and waiting time are found in some repeating FRBs \citep{Wang2017, Katz2018, Cheng2020, Hewitt2022, Jahns2023, Wang2023}. A large sample of FRB 121102 observed by the Five-hundred-meter Aperture Spherical radio Telescope \citep{Li2021a} also confirmed the power-law distributions of energy and waiting time \citep{Zhang2021}.

In this paper, we propose a capture model to explicate the glitch/anti-glitch and the associated FRB-like bursts in a unified way. 
In this model, both the glitch and the anti-glitch can be naturally explained. 
The FRB-like bursts are produced when the disrupted fragments penetrate through the stellar magnetosphere. 
The time interval between the glitch/anti-glitch and the FRB-like bursts can be interpreted as the fallback process of the bound material.  
Our main conclusions are as follows.

1. A iron-nickel asteroid is tidally disrupted by the magnetar at the tidal disruption radius $R_{\rm t}\sim 8.7\times 10^{10}\ \rm{cm}$. 
Then the angular momentum of the captured asteroid transfer to the magnetar at the magnetosphere radius ${R_{\rm m1}\sim 9.45\times 10^{9}}\ \rm{cm}$ for the anti-glitch and ${R_{\rm m2}\sim 4.25\times 10^{9}}\ \rm{cm}$ for the glitch, respectively. 
The lower-limit mass of the asteroid is estimated as ${m_1\sim 6.76\times 10^{21}}\ \rm{g}$ and ${m_2\sim 1.11\times10^{23}}\ \rm{g}$ for the anti-glitch and glitch, respectively. 
These results are consistent with the mass range of asteroids $1.0\times 10^{10}\ \rm{g} - 1.0 \times 10^{24}\ \rm{g}$ from observations.

2. 
The disrupted and bound asteroid falls back to the magnetar from the magnetosphere radius ${R_{\rm m}}$ and eventually is accreted onto the surface of the magnetar. The fallback time is consistent with the observed time delay between the glitch/anti-glitch and the radio burst.
In the case of anti-glitch, a train of fragments with collective mass $m_0\sim 1.7\times 10^{16}\ \rm{g}$ produces bright radio burst with energy $E_{\rm radio}\sim 1.86\times 10^{31}\ \rm{erg}$ at the interaction radius $R_{\rm I}\sim 3.05\times 10^7\ \rm{cm}$.

3. The luminosity of the possible associated X-ray burst is about $8.74\times 10^{35}\ \rm{erg\ s^{-1}}$, which is comparable to the 5$\sigma$ detection upper-limit of the X-ray luminosity. 
This is consistent with the non-detection of X-ray bursts during the anti-glitch and FRB-like bursts.  

4. The time interval between the glitch and the anti-glitch of SGR J1935+2154 is about 158 days. 
According to \cite{Dai2016}, the edge-on collision between the asteroid belt and the pulsar implies two travels through the belt, and the pulsar captures the asteroid twice resulting in glitch/anti-glitch. 
The time interval between them is $\sim R_{\rm a}/v_{\odot}\sim 170(R_{\rm a}/20\ \rm{au})(v_{\odot,7}/2)^{-1}$ days, where $R_{\rm a}\sim 20$ au is the inner radius of an asteroid belt \citep{DeMeo2014} and  $v_{\odot}\sim 2\times 10^7\ \rm{cm}\ \rm{s^{-1}}$  is the average proper velocity of the magnetar. The parameters used here are reasonable for the Solar System. 
This time interval is consistent with the observed time interval between the glitch and the anti-glitch. 
However, the rate of capture of an asteroid to produce the observed glitch/anti-glitch is difficult to determine, cause the angular momentum of the asteroid parallel/anti-parallel to the spinning magnetar is a rare event. 
Glitch/anti-glitch accompanied with bright radio bursts are so rare in observations.


\begin{table*}
\centering
\caption{Definitions of Some Parameters of the Capture Model}
\begin{threeparttable}
\begin{tabular}{ccc}
\hline
Description                                    &   Symbol       & Dependence   \\
\hline
Tidal disruption radius                        & $R_{\rm t}$    & $\sim 8.7\times 10^{10}\rho_{0.9}^{-1/3}\ \rm{cm}$ \\
Magnetosphere radius\tnote{a}                        & $R_{\rm m}$    & $\sim 9.45\times 10^{9}\mu_{32.5}^{2/3}\Delta t_{4.9}^{1/3}I_{45}^{-1/3}\Delta\nu_{1,-5.7}^{-1/3}  \ \rm{cm}$ \\
Interaction radius\tnote{a}                    & $R_{\rm I}$    & $ \sim3.05\times 10^{7}\rho_{ 0.9}^{-1/5}B_{14.5}^{2/5} \ \rm{cm}$ \\
Mass accretion rate\tnote{a}                   & $\dot{m}$      & $\sim 7.82\times 10^{16}I_{45}^{7/6}\Delta\nu_{1,-5.7}^{7/6}\mu_{32.5}^{-1/3}\Delta t_{4.9}^{-7/6}  \ \rm{g\ s^{-1}}$ \\
Mass of the asteroid\tnote{a}                  & $m$            & $\sim 6.76\times 10^{21}I_{45}^{7/6}\Delta\nu_{1,-5.7}^{7/6}\mu_{32.5}^{-1/3}\Delta t_{4.9}^{-1/6} \ \rm{g}$ \\
Mass of the heaviest broken fragment\tnote{a}  & $m_0$          & $\sim1.7\times 10^{16}\delta t_{-3.6}^{3} \rho_{0.9}  R_{\rm b,9.8}^{-3/2} \ \rm{g}$\\
Characteristic frequency of curvature radiation\tnote{a}              & $\nu_{\rm c}$  & $\sim465.49R_{\rm I, 7.5}^{11/4}\mu_{32.5}^{-3/2} \ \rm{MHz} $   \\
Luminosity of the FRB-like burst\tnote{a}         & $L_{\rm tot}$  & $\sim3.32\times 10^{32}R_{\rm I, 7.5}^{-27/4}\mu_{32.5}^{3/2} m_{0, 16.2}^{8/9}\rho_{ 0.9}^{-14/9} \ \rm{erg\ s^{-1}}$ \\
Luminosity of the X-ray burst\tnote{a}         & $L_{\rm X}$    & $\sim 8.74\times 10^{35}\eta_{-4.7}^{-1}L_{\rm tot, 32.5} \ \rm{erg\ s^{-1}}$ \\
\hline
\end{tabular}
\begin{tablenotes}
\footnotesize
\item[a] This calculation is for the case of the anti-glitch. 
\end{tablenotes}
\end{threeparttable}
\label{data}
\vspace{0.5cm}
\end{table*}

\section*{Acknowledgements}
We appreciate the referee for valuable comments and suggestions,
which have helped to improve this manuscript.
We thank Zigao Dai, Chen Deng, Guoqiang Zhang, Ziqian Hua, Yongfeng Huang, Jinjun Geng, and Zenan Liu for their helpful discussions. This work was supported by the National Natural Science Foundation of China (grant No. 12273009), the National SKA Program of
China (grant No. 2022SKA0130100), and the China Manned Spaced Project (CMS-CSST-2021-A12).

\section*{Data Availability}
No new data were generated in support of this theoretical research.

\bibliographystyle{mnras}
\bibliography{ms} 

\begin{thebibliography}{}
\makeatletter
\relax
\def\mn@urlcharsother{\let\do\@makeother \do\$\do\&\do\#\do\^\do\_\do\%\do\~}
\def\mn@doi{\begingroup\mn@urlcharsother \@ifnextchar [ {\mn@doi@}
  {\mn@doi@[]}}
\def\mn@doi@[#1]#2{\def\@tempa{#1}\ifx\@tempa\@empty \href
  {http://dx.doi.org/#2} {doi:#2}\else \href {http://dx.doi.org/#2} {#1}\fi
  \endgroup}
\def\mn@eprint#1#2{\mn@eprint@#1:#2::\@nil}
\def\mn@eprint@arXiv#1{\href {http://arxiv.org/abs/#1} {{\tt arXiv:#1}}}
\def\mn@eprint@dblp#1{\href {http://dblp.uni-trier.de/rec/bibtex/#1.xml}
  {dblp:#1}}
\def\mn@eprint@#1:#2:#3:#4\@nil{\def\@tempa {#1}\def\@tempb {#2}\def\@tempc
  {#3}\ifx \@tempc \@empty \let \@tempc \@tempb \let \@tempb \@tempa \fi \ifx
  \@tempb \@empty \def\@tempb {arXiv}\fi \@ifundefined
  {mn@eprint@\@tempb}{\@tempb:\@tempc}{\expandafter \expandafter \csname
  mn@eprint@\@tempb\endcsname \expandafter{\@tempc}}}

\bibitem[\protect\citeauthoryear{{Bochenek}, {Ravi}, {Belov}, {Hallinan},
  {Kocz}, {Kulkarni}  \& {McKenna}}{{Bochenek} et~al.}{2020}]{Bochenek2020}
{Bochenek} C.~D.,  {Ravi} V.,  {Belov} K.~V.,  {Hallinan} G.,  {Kocz} J.,
  {Kulkarni} S.~R.,   {McKenna} D.~L.,  2020, \mn@doi [\nat]
  {10.1038/s41586-020-2872-x}, \href
  {https://ui.adsabs.harvard.edu/abs/2020Natur.587...59B} {587, 59}

\bibitem[\protect\citeauthoryear{{Burnard}, {Arons}  \& {Lea}}{{Burnard}
  et~al.}{1983}]{Burnard1983}
{Burnard} D.~J.,  {Arons} J.,   {Lea} S.~M.,  1983, \mn@doi [\apj]
  {10.1086/160768}, \href
  {https://ui.adsabs.harvard.edu/abs/1983ApJ...266..175B} {266, 175}

\bibitem[\protect\citeauthoryear{{CHIME/FRB Collaboration} et~al.,}{{CHIME/FRB
  Collaboration} et~al.}{2020}]{CHIME2020}
{CHIME/FRB Collaboration} et~al., 2020, \mn@doi [\nat]
  {10.1038/s41586-020-2863-y}, \href
  {https://ui.adsabs.harvard.edu/abs/2020Natur.587...54C} {587, 54}

\bibitem[\protect\citeauthoryear{{Cannizzo}, {Lee}  \& {Goodman}}{{Cannizzo}
  et~al.}{1990}]{Cannizzo1990}
{Cannizzo} J.~K.,  {Lee} H.~M.,   {Goodman} J.,  1990, \mn@doi [\apj]
  {10.1086/168442}, \href
  {https://ui.adsabs.harvard.edu/abs/1990ApJ...351...38C} {351, 38}

\bibitem[\protect\citeauthoryear{{Carlin} \& {Melatos}}{{Carlin} \&
  {Melatos}}{2021}]{Carlin2021}
{Carlin} J.~B.,  {Melatos} A.,  2021, \mn@doi [\apj]
  {10.3847/1538-4357/ac06a2}, \href
  {https://ui.adsabs.harvard.edu/abs/2021ApJ...917....1C} {917, 1}

\bibitem[\protect\citeauthoryear{{Cheng}, {Zhang}  \& {Wang}}{{Cheng}
  et~al.}{2020}]{Cheng2020}
{Cheng} Y.,  {Zhang} G.~Q.,   {Wang} F.~Y.,  2020, \mn@doi [\mnras]
  {10.1093/mnras/stz3085}, \href
  {https://ui.adsabs.harvard.edu/abs/2020MNRAS.491.1498C} {491, 1498}

\bibitem[\protect\citeauthoryear{{Colgate} \& {Petschek}}{{Colgate} \&
  {Petschek}}{1981}]{Colgate1981}
{Colgate} S.~A.,  {Petschek} A.~G.,  1981, \mn@doi [\apj] {10.1086/159201},
  \href {https://ui.adsabs.harvard.edu/abs/1981ApJ...248..771C} {248, 771}

\bibitem[\protect\citeauthoryear{{Cordes} \& {Shannon}}{{Cordes} \&
  {Shannon}}{2008}]{Cordes2008}
{Cordes} J.~M.,  {Shannon} R.~M.,  2008, \mn@doi [\apj] {10.1086/589425}, \href
  {https://ui.adsabs.harvard.edu/abs/2008ApJ...682.1152C} {682, 1152}

\bibitem[\protect\citeauthoryear{{Dai}}{{Dai}}{2020}]{Dai2020}
{Dai} Z.~G.,  2020, \mn@doi [\apjl] {10.3847/2041-8213/aba11b}, \href
  {https://ui.adsabs.harvard.edu/abs/2020ApJ...897L..40D} {897, L40}

\bibitem[\protect\citeauthoryear{{Dai}, {Wang}, {Wu}  \& {Huang}}{{Dai}
  et~al.}{2016}]{Dai2016}
{Dai} Z.~G.,  {Wang} J.~S.,  {Wu} X.~F.,   {Huang} Y.~F.,  2016, \mn@doi [\apj]
  {10.3847/0004-637X/829/1/27}, \href
  {https://ui.adsabs.harvard.edu/abs/2016ApJ...829...27D} {829, 27}

\bibitem[\protect\citeauthoryear{{DeMeo} \& {Carry}}{{DeMeo} \&
  {Carry}}{2014}]{DeMeo2014}
{DeMeo} F.~E.,  {Carry} B.,  2014, \mn@doi [\nat] {10.1038/nature12908}, \href
  {https://ui.adsabs.harvard.edu/abs/2014Natur.505..629D} {505, 629}

\bibitem[\protect\citeauthoryear{{Du}}{{Du}}{2023}]{Du2023}
{Du} S.,  2023, arXiv e-prints, \href
  {https://ui.adsabs.harvard.edu/abs/arXiv:2301.04602} {p. arXiv:2301.04602}

\bibitem[\protect\citeauthoryear{{Ge} et~al.,}{{Ge} et~al.}{2022}]{Ge2022}
{Ge} M.,  et~al., 2022, arXiv e-prints, \href
  {https://ui.adsabs.harvard.edu/abs/2022arXiv221103246G} {p. arXiv:2211.03246}

\bibitem[\protect\citeauthoryear{{Geng} \& {Huang}}{{Geng} \&
  {Huang}}{2015}]{Geng2015}
{Geng} J.~J.,  {Huang} Y.~F.,  2015, \mn@doi [\apj]
  {10.1088/0004-637X/809/1/24}, \href
  {https://ui.adsabs.harvard.edu/abs/2015ApJ...809...24G} {809, 24}

\bibitem[\protect\citeauthoryear{{Geng}, {Li}, {Li}, {Xiong}, {Kuiper}  \&
  {Huang}}{{Geng} et~al.}{2020}]{Geng2020}
{Geng} J.-J.,  {Li} B.,  {Li} L.-B.,  {Xiong} S.-L.,  {Kuiper} R.,   {Huang}
  Y.-F.,  2020, \mn@doi [\apjl] {10.3847/2041-8213/aba83c}, \href
  {https://ui.adsabs.harvard.edu/abs/2020ApJ...898L..55G} {898, L55}

\bibitem[\protect\citeauthoryear{{Ghosh} \& {Lamb}}{{Ghosh} \&
  {Lamb}}{1979}]{Ghosh1979a}
{Ghosh} P.,  {Lamb} F.~K.,  1979, \mn@doi [\apj] {10.1086/157285}, \href
  {https://ui.adsabs.harvard.edu/abs/1979ApJ...232..259G} {232, 259}

\bibitem[\protect\citeauthoryear{{Good} \& {Chime/Frb Collaboration}}{{Good} \&
  {Chime/Frb Collaboration}}{2020}]{Good2020}
{Good} D.,  {Chime/Frb Collaboration} 2020, The Astronomer's Telegram, \href
  {https://ui.adsabs.harvard.edu/abs/2020ATel14074....1G} {14074, 1}

\bibitem[\protect\citeauthoryear{{Hameury}, {King}  \& {Lasota}}{{Hameury}
  et~al.}{1986}]{Hameury1986}
{Hameury} J.~M.,  {King} A.~R.,   {Lasota} J.~P.,  1986, \mn@doi [\mnras]
  {10.1093/mnras/218.4.695}, \href
  {https://ui.adsabs.harvard.edu/abs/1986MNRAS.218..695H} {218, 695}

\bibitem[\protect\citeauthoryear{{Haskell} \& {Melatos}}{{Haskell} \&
  {Melatos}}{2015}]{Haskell2015}
{Haskell} B.,  {Melatos} A.,  2015, \mn@doi [International Journal of Modern
  Physics D] {10.1142/S0218271815300086}, \href
  {https://ui.adsabs.harvard.edu/abs/2015IJMPD..2430008H} {24, 1530008}

\bibitem[\protect\citeauthoryear{{Hewitt} et~al.,}{{Hewitt}
  et~al.}{2022}]{Hewitt2022}
{Hewitt} D.~M.,  et~al., 2022, \mn@doi [\mnras] {10.1093/mnras/stac1960}, \href
  {https://ui.adsabs.harvard.edu/abs/2022MNRAS.515.3577H} {515, 3577}

\bibitem[\protect\citeauthoryear{{Hills}}{{Hills}}{1975}]{Hills1975}
{Hills} J.~G.,  1975, \mn@doi [\nat] {10.1038/254295a0}, \href
  {https://ui.adsabs.harvard.edu/abs/1975Natur.254..295H} {254, 295}

\bibitem[\protect\citeauthoryear{{Hooker}, {Newton}  \& {Li}}{{Hooker}
  et~al.}{2015}]{Hooker2015}
{Hooker} J.,  {Newton} W.~G.,   {Li} B.-A.,  2015, \mn@doi [\mnras]
  {10.1093/mnras/stv582}, \href
  {https://ui.adsabs.harvard.edu/abs/2015MNRAS.449.3559H} {449, 3559}

\bibitem[\protect\citeauthoryear{{Huang} \& {Geng}}{{Huang} \&
  {Geng}}{2014}]{Huang2014}
{Huang} Y.~F.,  {Geng} J.~J.,  2014, \mn@doi [\apjl]
  {10.1088/2041-8205/782/2/L20}, \href
  {https://ui.adsabs.harvard.edu/abs/2014ApJ...782L..20H} {782, L20}

\bibitem[\protect\citeauthoryear{{Israel} et~al.,}{{Israel}
  et~al.}{2016}]{Israel2016}
{Israel} G.~L.,  et~al., 2016, \mn@doi [\mnras] {10.1093/mnras/stw008}, \href
  {https://ui.adsabs.harvard.edu/abs/2016MNRAS.457.3448I} {457, 3448}

\bibitem[\protect\citeauthoryear{{Jahns} et~al.,}{{Jahns}
  et~al.}{2023}]{Jahns2023}
{Jahns} J.~N.,  et~al., 2023, \mn@doi [\mnras] {10.1093/mnras/stac3446}, \href
  {https://ui.adsabs.harvard.edu/abs/2023MNRAS.519..666J} {519, 666}

\bibitem[\protect\citeauthoryear{{Kantor} \& {Gusakov}}{{Kantor} \&
  {Gusakov}}{2014}]{Kantor2014}
{Kantor} E.~M.,  {Gusakov} M.~E.,  2014, \mn@doi [\apjl]
  {10.1088/2041-8205/797/1/L4}, \href
  {https://ui.adsabs.harvard.edu/abs/2014ApJ...797L...4K} {797, L4}

\bibitem[\protect\citeauthoryear{{Katz}}{{Katz}}{2014}]{Katz2014}
{Katz} J.~I.,  2014, \mn@doi [\apss] {10.1007/s10509-013-1732-7}, \href
  {https://ui.adsabs.harvard.edu/abs/2014Ap&SS.349..611K} {349, 611}

\bibitem[\protect\citeauthoryear{{Katz}}{{Katz}}{2018}]{Katz2018}
{Katz} J.~I.,  2018, \mn@doi [\mnras] {10.1093/mnras/sty366}, \href
  {https://ui.adsabs.harvard.edu/abs/2018MNRAS.476.1849K} {476, 1849}

\bibitem[\protect\citeauthoryear{{Lamb}, {Pethick}  \& {Pines}}{{Lamb}
  et~al.}{1973}]{Lamb1973}
{Lamb} F.~K.,  {Pethick} C.~J.,   {Pines} D.,  1973, \mn@doi [\apj]
  {10.1086/152325}, \href
  {https://ui.adsabs.harvard.edu/abs/1973ApJ...184..271L} {184, 271}

\bibitem[\protect\citeauthoryear{{Li} et~al.,}{{Li} et~al.}{2021a}]{Li2021}
{Li} C.~K.,  et~al., 2021a, \mn@doi [Nature Astronomy]
  {10.1038/s41550-021-01302-6}, \href
  {https://ui.adsabs.harvard.edu/abs/2021NatAs...5..378L} {5, 378}

\bibitem[\protect\citeauthoryear{{Li} et~al.,}{{Li} et~al.}{2021b}]{Li2021a}
{Li} D.,  et~al., 2021b, \mn@doi [\nat] {10.1038/s41586-021-03878-5}, \href
  {https://ui.adsabs.harvard.edu/abs/2021Natur.598..267L} {598, 267}

\bibitem[\protect\citeauthoryear{{Litwin} \& {Rosner}}{{Litwin} \&
  {Rosner}}{2001}]{Litwin2001}
{Litwin} C.,  {Rosner} R.,  2001, \mn@doi [\prl] {10.1103/PhysRevLett.86.4745},
  \href {https://ui.adsabs.harvard.edu/abs/2001PhRvL..86.4745L} {86, 4745}

\bibitem[\protect\citeauthoryear{{Lu}, {Kumar}  \& {Zhang}}{{Lu}
  et~al.}{2020}]{Lu2020}
{Lu} W.,  {Kumar} P.,   {Zhang} B.,  2020, \mn@doi [\mnras]
  {10.1093/mnras/staa2450}, \href
  {https://ui.adsabs.harvard.edu/abs/2020MNRAS.498.1397L} {498, 1397}

\bibitem[\protect\citeauthoryear{{Lyutikov}}{{Lyutikov}}{2013}]{Lyutikov2013}
{Lyutikov} M.,  2013, arXiv e-prints, \href
  {https://ui.adsabs.harvard.edu/abs/2013arXiv1306.2264L} {p. arXiv:1306.2264}

\bibitem[\protect\citeauthoryear{{Margalit}, {Metzger}  \& {Sironi}}{{Margalit}
  et~al.}{2020}]{Margalit2020}
{Margalit} B.,  {Metzger} B.~D.,   {Sironi} L.,  2020, \mn@doi [\mnras]
  {10.1093/mnras/staa1036}, \href
  {https://ui.adsabs.harvard.edu/abs/2020MNRAS.494.4627M} {494, 4627}

\bibitem[\protect\citeauthoryear{{Melatos}, {Peralta}  \& {Wyithe}}{{Melatos}
  et~al.}{2008}]{Melatos2008}
{Melatos} A.,  {Peralta} C.,   {Wyithe} J.~S.~B.,  2008, \mn@doi [\apj]
  {10.1086/523349}, \href
  {https://ui.adsabs.harvard.edu/abs/2008ApJ...672.1103M} {672, 1103}

\bibitem[\protect\citeauthoryear{{Mereghetti} et~al.,}{{Mereghetti}
  et~al.}{2020}]{Mereghetti2020}
{Mereghetti} S.,  et~al., 2020, \mn@doi [\apjl] {10.3847/2041-8213/aba2cf},
  \href {https://ui.adsabs.harvard.edu/abs/2020ApJ...898L..29M} {898, L29}

\bibitem[\protect\citeauthoryear{{Metzger}, {Beniamini}  \&
  {Giannios}}{{Metzger} et~al.}{2018}]{Metzger2018}
{Metzger} B.~D.,  {Beniamini} P.,   {Giannios} D.,  2018, \mn@doi [\apj]
  {10.3847/1538-4357/aab70c}, \href
  {https://ui.adsabs.harvard.edu/abs/2018ApJ...857...95M} {857, 95}

\bibitem[\protect\citeauthoryear{{Morley} \& {Garcia-Pelayo}}{{Morley} \&
  {Garcia-Pelayo}}{1993}]{Morley1993}
{Morley} P.~D.,  {Garcia-Pelayo} R.,  1993, \mn@doi [EPL (Europhysics Letters)]
  {10.1209/0295-5075/23/3/005}, \href
  {https://ui.adsabs.harvard.edu/abs/1993EL.....23..185M} {23, 185}

\bibitem[\protect\citeauthoryear{{Perets}, {Li}, {Lombardi}  \&
  {Milcarek}}{{Perets} et~al.}{2016}]{Perets2016}
{Perets} H.~B.,  {Li} Z.,  {Lombardi} James~C. J.,   {Milcarek} Stephen~R. J.,
  2016, \mn@doi [\apj] {10.3847/0004-637X/823/2/113}, \href
  {https://ui.adsabs.harvard.edu/abs/2016ApJ...823..113P} {823, 113}

\bibitem[\protect\citeauthoryear{{Pizzochero}}{{Pizzochero}}{2011}]{Pizzochero2011}
{Pizzochero} P.~M.,  2011, \mn@doi [\apjl] {10.1088/2041-8205/743/1/L20}, \href
  {https://ui.adsabs.harvard.edu/abs/2011ApJ...743L..20P} {743, L20}

\bibitem[\protect\citeauthoryear{{Rees}}{{Rees}}{1988}]{Rees1988}
{Rees} M.~J.,  1988, \mn@doi [\nat] {10.1038/333523a0}, \href
  {https://ui.adsabs.harvard.edu/abs/1988Natur.333..523R} {333, 523}

\bibitem[\protect\citeauthoryear{{Ridnaia} et~al.,}{{Ridnaia}
  et~al.}{2021}]{Ridnaia2021}
{Ridnaia} A.,  et~al., 2021, \mn@doi [Nature Astronomy]
  {10.1038/s41550-020-01265-0}, \href
  {https://ui.adsabs.harvard.edu/abs/2021NatAs...5..372R} {5, 372}

\bibitem[\protect\citeauthoryear{{Shen}, {Zou}, {Yang}, {Zheng}  \&
  {Wang}}{{Shen} et~al.}{2023}]{Shen2023}
{Shen} J.-Y.,  {Zou} Y.-C.,  {Yang} S.-H.,  {Zheng} X.-P.,   {Wang} K.,  2023,
  \mn@doi [arXiv e-prints] {10.48550/arXiv.2304.10871}, \href
  {https://ui.adsabs.harvard.edu/abs/2023arXiv230410871S} {p. arXiv:2304.10871}

\bibitem[\protect\citeauthoryear{{Tavani} et~al.,}{{Tavani}
  et~al.}{2021}]{Tavani2021}
{Tavani} M.,  et~al., 2021, \mn@doi [Nature Astronomy]
  {10.1038/s41550-020-01276-x}, \href
  {https://ui.adsabs.harvard.edu/abs/2021NatAs...5..401T} {5, 401}

\bibitem[\protect\citeauthoryear{{Thompson}, {Duncan}, {Woods}, {Kouveliotou},
  {Finger}  \& {van Paradijs}}{{Thompson} et~al.}{2000}]{Thompson2000}
{Thompson} C.,  {Duncan} R.~C.,  {Woods} P.~M.,  {Kouveliotou} C.,  {Finger}
  M.~H.,   {van Paradijs} J.,  2000, \mn@doi [\apj] {10.1086/317072}, \href
  {https://ui.adsabs.harvard.edu/abs/2000ApJ...543..340T} {543, 340}

\bibitem[\protect\citeauthoryear{{Tong}}{{Tong}}{2014}]{Tong2014}
{Tong} H.,  2014, \mn@doi [\apj] {10.1088/0004-637X/784/2/86}, \href
  {https://ui.adsabs.harvard.edu/abs/2014ApJ...784...86T} {784, 86}

\bibitem[\protect\citeauthoryear{{Tong}}{{Tong}}{2023}]{Tong2022}
{Tong} H.,  2023, \mn@doi [Research in Astronomy and Astrophysics]
  {10.1088/1674-4527/acae70}, \href
  {https://ui.adsabs.harvard.edu/abs/2023RAA....23b5013T} {23, 025013}

\bibitem[\protect\citeauthoryear{{Wang} \& {Yu}}{{Wang} \&
  {Yu}}{2017}]{Wang2017}
{Wang} F.~Y.,  {Yu} H.,  2017, \mn@doi [Journal of Cosmology and Astroparticle
  Physics] {10.1088/1475-7516/2017/03/023}, \href
  {https://ui.adsabs.harvard.edu/abs/2017JCAP...03..023W} {2017, 023}

\bibitem[\protect\citeauthoryear{{Wang}, {Xu}, {Wang}, {Du}, {Cheng}, {Zheng}
  \& {Xu}}{{Wang} et~al.}{2021}]{WangWH2021}
{Wang} W.-H.,  {Xu} H.,  {Wang} W.-Y.,  {Du} S.,  {Cheng} Q.,  {Zheng} X.-P.,
  {Xu} R.-X.,  2021, \mn@doi [\mnras] {10.1093/mnras/stab2213}, \href
  {https://ui.adsabs.harvard.edu/abs/2021MNRAS.507.2208W} {507, 2208}

\bibitem[\protect\citeauthoryear{{Wang}, {Ge}, {Huang}  \& {Zheng}}{{Wang}
  et~al.}{2022a}]{WangWH2022}
{Wang} W.-H.,  {Ge} M.-Y.,  {Huang} X.,   {Zheng} X.-P.,  2022a, arXiv
  e-prints, \href {https://ui.adsabs.harvard.edu/abs/2022arXiv221108151W} {p.
  arXiv:2211.08151}

\bibitem[\protect\citeauthoryear{{Wang}, {Zhang}, {Dai}  \& {Cheng}}{{Wang}
  et~al.}{2022b}]{WangF2022}
{Wang} F.~Y.,  {Zhang} G.~Q.,  {Dai} Z.~G.,   {Cheng} K.~S.,  2022b, \mn@doi
  [Nature Communications] {10.1038/s41467-022-31923-y}, \href
  {https://ui.adsabs.harvard.edu/abs/2022NatCo..13.4382W} {13, 4382}

\bibitem[\protect\citeauthoryear{{Wang}, {Wu}  \& {Dai}}{{Wang}
  et~al.}{2023}]{Wang2023}
{Wang} F.~Y.,  {Wu} Q.,   {Dai} Z.~G.,  2023, \mn@doi [arXiv e-prints]
  {10.48550/arXiv.2302.06802}, \href
  {https://ui.adsabs.harvard.edu/abs/2023arXiv230206802W} {p. arXiv:2302.06802}

\bibitem[\protect\citeauthoryear{{Wu}, {Zhang}, {Wang}  \& {Dai}}{{Wu}
  et~al.}{2020}]{Wu2020}
{Wu} Q.,  {Zhang} G.~Q.,  {Wang} F.~Y.,   {Dai} Z.~G.,  2020, \mn@doi [\apjl]
  {10.3847/2041-8213/abaef1}, \href
  {https://ui.adsabs.harvard.edu/abs/2020ApJ...900L..26W} {900, L26}

\bibitem[\protect\citeauthoryear{{Yang} \& {Zhang}}{{Yang} \&
  {Zhang}}{2018}]{Yang2018}
{Yang} Y.-P.,  {Zhang} B.,  2018, \mn@doi [\apj] {10.3847/1538-4357/aae685},
  \href {https://ui.adsabs.harvard.edu/abs/2018ApJ...868...31Y} {868, 31}

\bibitem[\protect\citeauthoryear{{Younes} et~al.,}{{Younes}
  et~al.}{2023}]{Younes2022}
{Younes} G.,  et~al., 2023, \mn@doi [Nature Astronomy]
  {10.1038/s41550-022-01865-y}, \href
  {https://ui.adsabs.harvard.edu/abs/2023NatAs...7..339Y} {7, 339}

\bibitem[\protect\citeauthoryear{{Zhang}, {Wang}, {Wu}, {Wang}, {Li}, {Dai}  \&
  {Zhang}}{{Zhang} et~al.}{2021}]{Zhang2021}
{Zhang} G.~Q.,  {Wang} P.,  {Wu} Q.,  {Wang} F.~Y.,  {Li} D.,  {Dai} Z.~G.,
  {Zhang} B.,  2021, \mn@doi [\apjl] {10.3847/2041-8213/ac2a3b}, \href
  {https://ui.adsabs.harvard.edu/abs/2021ApJ...920L..23Z} {920, L23}

\bibitem[\protect\citeauthoryear{{Zhong}, {Dai}, {Zhang}  \& {Deng}}{{Zhong}
  et~al.}{2020}]{Zhong2020}
{Zhong} S.-Q.,  {Dai} Z.-G.,  {Zhang} H.-M.,   {Deng} C.-M.,  2020, \mn@doi
  [\apjl] {10.3847/2041-8213/aba262}, \href
  {https://ui.adsabs.harvard.edu/abs/2020ApJ...898L...5Z} {898, L5}

\bibitem[\protect\citeauthoryear{{Zhu} et~al.,}{{Zhu} et~al.}{2020}]{Zhu2020}
{Zhu} W.,  et~al., 2020, The Astronomer's Telegram, \href
  {https://ui.adsabs.harvard.edu/abs/2020ATel14084....1Z} {14084, 1}

\bibitem[\protect\citeauthoryear{{Zubovas}, {Nayakshin}  \&
  {Markoff}}{{Zubovas} et~al.}{2012}]{Zubovas2012}
{Zubovas} K.,  {Nayakshin} S.,   {Markoff} S.,  2012, \mn@doi [\mnras]
  {10.1111/j.1365-2966.2011.20389.x}, \href
  {https://ui.adsabs.harvard.edu/abs/2012MNRAS.421.1315Z} {421, 1315}

\makeatother
\end{thebibliography}

\bsp	
\label{lastpage}
\end{document}